# High Performance Semiconducting Enriched Carbon Nanotube Thin Film Transistors Using Metallic Carbon Nanotube Electrode

Biddut K. Sarker,[1,2,†] Narae Kang,[1,2,†] and Saiful I. Khondaker[1,2,3] *

[1]Nanoscience Technology Center, [2]Department of Physics, [3]School of Electrical Engineering and Computer Science, University of Central Florida, 12424 Research Parkway, Suite 400, Orlando, Florida 32826, USA

[†] These authors contributed equally to this work
* To whom correspondence should be addressed. E-mail: saiful@ucf.edu

## Abstract

High-performance solution-processed short-channel carbon nanotube (CNT) thin film transistors (TFTs) are fabricated using densely aligned arrays of metallic CNTs (m-CNTs) as source and drain electrodes, and aligned arrays of semiconducting enriched CNTs (s-CNTs) as channel material. The electrical transport measurements at room temperature show that the m-CNT contacted s-CNT array devices with a 2 μm channel length perform superiorly to those of control Pd contacted s-CNT devices. The m-CNT contacted devices exhibit a maximum (average) on-conductance of 36.5 μS (19.2 μS), transconductance of 2.6 μS (1.2 μS), mobility of 51 $cm^2$/Vs (25 $cm^2$/Vs), and current on-off ratio of $1.1\times10^6$ ($2.5\times10^5$). These values are almost an order of magnitude higher than that of control Pd contacted devices with the same channel length and s-CNT linear density. The low temperature charge transport measurements suggest that these improved performances are due to the lower charge injection barrier of m-CNT/s-CNT array devices compared to Pd/s-CNT array devices. We attribute the lower injection barrier to unique geometry of our devices. In addition to using semiconducting enriched CNT, our results suggest that using metallic CNT as an electrode can significantly enhance the performance of CNT TFTs.

## 1. Introduction

Because of the exceptional electronic and mechanical properties of carbon nanotubes (CNTs), thin film transistors (TFTs) fabricated with CNTs have attracted a great deal of attention as promising components of the next-generation of flexible and transparent electronic devices, sensors and high frequency devices.[1-11] The TFTs fabricated from a network of CNTs can be advantageous to individual nanotube devices as they provide more device to device homogeneity and cover large areas. Since in the CNT TFTs a large number of nanotubes contribute in the transportation of charge simultaneously, the output current can also be significantly increased. Several recent studies show that the performance of the CNT TFT depends on the channel length of the TFTs, and the content of semiconducting *versus* metallic CNTs.[8-18] The CNT TFTs have been fabricated using mixed CNTs either in a random network or in an aligned array.[10-13, 19-21] The mixed nanotubes grown by chemical vapor deposition (CVD) contain generally a large portion of metallic CNTs (m-CNTs). The metallic nanotube pathways tend to dominate the transport in the TFTs fabricated using mixed CNTs, which resulted in lower current on-off ratio. In order to increase the current on-off ratio, selective removal of metallic CNTs via electrical breakdown has been used,[12, 20-21] however, this method has detrimental effects on the remaining nanotubes in the network.[21-22]

Therefore, since only semiconducting CNTs (s-CNT) have application in FETs, it is important to use all semiconducting nanotubes for fabricating CNT TFTs.

Studies have shown that solution based sorting techniques, such as density gradient ultracentrifugation, can provide highly enriched s-CNTs in aqueous solution.[23] A few studies reported the fabrication of TFTs using highly pure s-CNTs either in a random network or in an aligned array.[8, 14-18, 23-25] In all of these studies, metal electrodes (Au, Pd) were used to fabricate TFTs. It was found that for the TFTs fabricated with highly enriched ( > 98% purity) s-CNT show better device performance (mainly mobility, and current on-off ratio) when the channel length of the transistors is much larger than average length of the nanotubes.[2, 16, 18] However, the performance of the TFTs decreases with decreasing the channel length.[2, 14, 16, 18] For these devices, s-CNT solution contains a very small fraction (<2%) of m-CNTs, and when the channel length of the device decreases, there is an increased probability (if the s-CNT density in the array is high) that the m-CNTs will form a percolating path which can reduce the current on-off ratio to less than 100.[14, 15] Although the decrease of current on-off ratio with decreasing channel length can be explained by the presence of m-CNTs, however, it is not clear why the mobility decreases with decreasing the channel length. A previous study showed that the mobility for 95% pure s-CNT devices is higher than 98% pure s-CNT devices.[16] In addition, the TFTs fabricated from mixed CNTs (1/3 metallic) show that the mobility decreases when m-CNTs are removed by electrical breakdown.[12, 19] These observations clearly indicate that the presence of m-CNTs increases the mobility. Thus, one can expect that the mobility of the s-CNT devices should be increased with decreasing channel length if the small fraction of m-CNT makes a percolating path between the electrodes. Therefore, the decrease of mobility with decreasing channel length for the highly enriched s-CNT TFT devices can not be explained by the percolation of small fraction of m-CNTs. In addition, for s-CNT TFTs where the on-off ratio was greater than 100 (meaning no m-CNT percolation), the decrease of mobility with decreasing channel length was still observed.[16] Burke group speculated that the decrease of mobility with decreasing the channel length may be due to the effect of contact. [18] In the long channel CNT TFTs, the contact resistance may be negligible in comparison to the channel resistance and hence the effect of contact resistance on the performance of long channel devices is small. In contrast, in the short channel devices, the contact resistance can become a significant fraction of the total resistance of the devices due to a reduction of channel resistance resulting in a reduction of mobility. The effect of contact is rather severe in short channel devices where the typical mobility is 2-15 $cm^2/Vs$ for a current on-off ratio of greater than 100.[14-16, 26] It should be noted that short channel CNT TFTs are important for high frequency applications as the on-current is increased with decreasing the channel length of the transistor and the cutoff frequency is inversely proportional to the square of the channel length.[2, 18] Enhancing the performance of short channel TFTs are also of great significance for realizing the overreaching goals of CNT based TFTs.

One possible way to improve the performance of short channel solution processed CNT TFTs could be to use aligned m-CNTs as electrode material and aligned s-CNTs as channel material where both m-CNTs and s-CNTs are longitudinal to each other. The m-CNTs with open-ended and parallel tips can enhance charge injection at m-CNT/s-CNTs interface due to electric field enhancement at these tips.[27, 28] In addition, longitudinal arrangement can also result in finite overlap of some m-CNT/s-CNT which will also enhance charge injection. Because of the higher charge injection from the m-CNTs, we can expect that the interfacial barrier at the m-CNT/s-CNT interface will be reduced which will improve the performance of the CNT TFTs.

In this paper, we show such a longitudinal arrangement between m-CNTs electrode and s-CNT channel and demonstrate that high performance short channel CNT TFTs can be realized from such an arrangement. Aligned arrays of m-CNTs as electrode with channel length (*L*) of 2 μm and aligned arrays of s-CNTs as channel materials were assembled via dielectrophoresis (DEP) from high quality aqueous solutions. From the electrical transport measurements of 13 m-CNT contacted s-CNT devices with a linear density of 1 s-CNT/μm at room temperature, we found a maximum (average) on-conductance of 36.5 μS (19.2 μS), transconductance of 2.6 μS (1.2 μS), mobility of 51 cm$^2$/Vs (25 cm$^2$/Vs), and current on-off ratio of $1.1\times10^6$ ($2.5\times10^5$). These values are almost an order of magnitude higher than those of the control devices of the same density and channel length, fabricated with Pd electrodes. In order to investigate the reason for improved device performance of our m-CNT contacted s-CNT devices, we carried out temperature dependent electron transport investigations. We found that our data can be modeled using the Schottky barrier model. The charge injection barrier for the m-CNT contacted s-CNT devices was found to be 25 meV, which is significantly lower than the barrier for Pd contacted s-CNT devices (83 meV) of the same s-CNT density, suggesting that the improved performance for m-CNT/s-CNT devices is a result of improved charge injection. Our results suggest that in addition to using semiconducting enriched CNT in the channel, using metallic CNT as the electrode can significantly enhance the performance of CNT TFTs.

## 2. Experimental Section

*The Fabrication of m-CNT Electrodes:* The high-quality, stable, and surfactant-free CNT aqueous solution was used to fabricate the CNT source and drain electrodes. The CNT aqueous solution was obtained from a commercial source, Brewer Science Inc., and mostly contains individual nanotubes with an average diameter and length of 1.7 nm and 1.5 μm, respectively (confirmed by scanning electron microscopy (SEM) and atomic force microscopy (AFM)).[29] The original concentration of the CNT solution was ~50μg/mL and was diluted to about a third of this value with di-ionized (DI) water for the assembly. The Pd patterns of 5 μm × 25 μm were used to align the CNT in a dense array via DEP. The Pd patterns were fabricated on a Si substrate with a thermally grown 250 nm thick SiO$_2$ layer using standard lithography process followed by e-beam evaporation of 30 nm Pd and lift-off. To assemble the array using DEP, a small drop of CNT solution (3 μl) was dropped on the chip containing the Pd patterns and an AC voltage of 5V$_{pp}$ with a frequency of 1MHz was applied for 30 seconds. The AC voltage creates a time averaged DEP force and align the CNTs in the direction of the electric field between the Pd patterns.[29] After the CNT assembly, a layer of PMMA was spin-coated at 4000 rpm for 60 seconds then baked at 180˚C for 15 min on a hot plate. Then, the CNT source and drain electrodes (with a channel length of 2 μm and a width 25 μm) were defined using EBL followed by precisely cutting the CNTs by plasma etching. Finally, the chip containing the CNT electrodes was kept in chloroform for 12 hrs to remove the remaining PMMA and afterwards, rinsed with acetone, IPA, and DI water and blown with nitrogen gas. More details of the CNT assembly and m-CNT electrode fabrication can be found in our previous works.[30-31]

*The Fabrication of s-CNT TFTs with m-CNT Electrodes:* Highly enriched (99%) s-CNT solution, purchased from Nanointegris, was used to fabricate the transistors for both m-CNT and Pd electrodes. The diameter and length of the s-CNTs were characterized by SEM and AFM. The diameter of the s-CNTs varied from 0.5 to 3.9 nm with an average of 1.7 nm, and the length of the s-CNTs varied from 0.7 to 4.0 μm with an average of 1.8 μm.[15] The concentration of original s-CNT solution was ~10μg/mL, which was then diluted by DI water to a solution concentration of 2 ng/ml. The DEP method was used again to assemble the s-CNTs in between the CNT electrodes by applying an AC voltage of 5 V$_{pp}$ at 1 MHz for 30 seconds.

*The Fabrication of s-CNT TFTs with Pd Electrodes (Control devices)*: We fabricated control Pd contaced s-CNT devices *via* DEP assembly of highly enriched (99%) s-CNTs following our previously reported technique.[15] The Pd electrodes deposited on Si/SiO$_2$ substrate had a channel length of 2 µm and width of 25 µm, exactly the same dimension of the m-CNT electrodes. The parameters for DEP assembly of the s-CNTs using Pd electrodes is the same as what has been used for m-CNTs electrodes. Both the m-CNT contacted and Pd contacted s-CNT devices were annealed at 200 $^0$C for one hour in an argon/hydrogen (Ar/H$_2$) environment.

*The Characterization of the Devices:* The SEM and AFM images of the devices were taken by the Zeiss Ultra-55 SEM and the Dimension 3100 AFM (Veeco). The electronic transport measurements at both room and low temperatures were taken using DL instruments 1211 current preamplifier and Keithley 2400 source meter interfaced with LabView program. To perform the temperature dependent transport measurements, the devices were bonded in chip carriers and loaded into the cryostat. The temperature was controlled by the Lakeshore temperature controller. Before taking any data, we waited 10 minutes for the temperature to stabilize and between successive measurements, we waited 15 minutes.

## 3. Results and Discussion

A schematic diagram of the fabrication steps of the s-CNT aligned array TFTs with aligned array m-CNT electrodes is shown in Figure 1(a). Figure 1(b) shows a SEM image of a part of the CNT aligned array. This array has a linear density of ~ 30 CNT/µm and a resistance of 550 Ω (see Figure S1a in supporting information) which corresponds to a sheet resistance of 2.75 kΩ/sq. Although we used mixed CNTs for this assembly, it has been shown that the DEP process favors m-CNTs over s-CNTs.[32] This is also evident by the low sheet resistance. In addition, the densely aligned CNT arrays do not show any gate-voltage dependence (Figure S1b), further implying the metallic behavior of the CNTs arrays. The low sheet resistance and absence of gate dependence confirms the metallic behavior making them ideal for electrode fabrication. Since we only used dense arrays for fabricating the CNT electrodes and they show the metallic behavior, we call the dense CNT arrays as m-CNT array. After the assembly, we opened an window of channel length 2 µm and width 25 µm on the m-CNTs arrays via electron beam lithography (EBL) and etched away the exposed area using oxygen plasma.[30,31] Figure 1(c) shows the SEM image and Figure 1(d) shows the AFM image of a part of a m-CNT electrode. The AFM image clearly shows that the CNTs in the electrode have open-ended tips and they are parallel to each other.

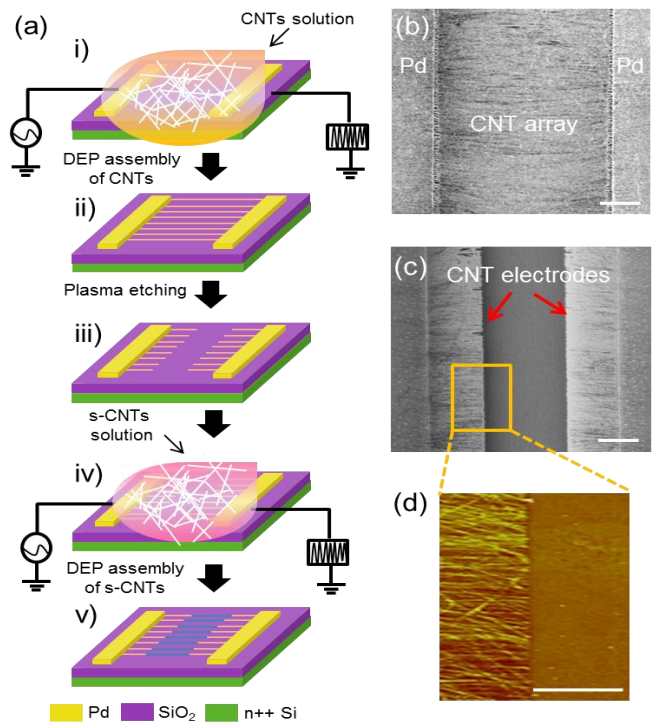

**Figure 1.** (a) Schematic diagram of fabrication steps for s-CNT TFTs with CNT electrodes. (i) DEP assembly of CNTs. (ii) Aligned CNT arrays *via* DEP between palladium (Pd) patterns. (iii) Opening of a window on the CNTs array *via* EBL and plasma etching. (iv) DEP assembly of s-CNTs in between CNT electrodes. (v) Schematic of s-CNTs TFT with CNT electrodes. SEM images of (b) CNT aligned array, (c) CNT electrode (after plasma etching). The channel length of the CNT electrode is 2 µm, and (d) AFM image of a part of the CNT electrode showing open ended and parallel tips. The scale bar indicates 1µm.

The s-CNT array was then assembled between the m-CNT electrodes as seen in Figure 2(a) showing an SEM image of a typical m-CNT/s-CNT device with a linear density of ~1 s-CNT/μm in the channel. Figure 2(b) shows a high magnification SEM image of the interface between m-CNT and s-CNT. This image demonstrates unique longitudinal arrangement between m-CNT electrode and s-CNT in the channel with a possible end-contact configuration where the m-CNTs and s-CNTs are connected in a head-to-head arrangement without any gap. All the devices reported in this study have the same linear density of s-CNT. For comparison, we also fabricated control Pd contacted s-CNT devices of same density and the same channel length and width.

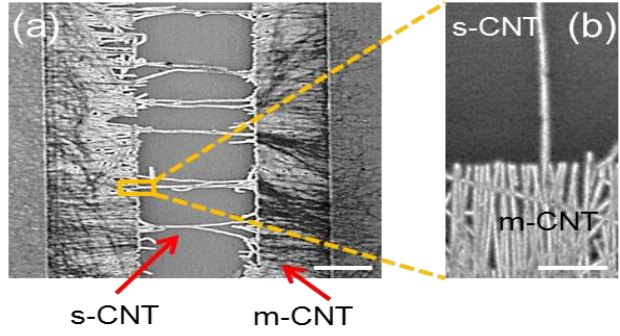

**Figure 2.** (a) SEM images of aligned array s-CNT with linear density of 1 s-CNT/μm between m-CNT electrodes with a channel length of 2 μm. (b) High magnification SEM image showing longitudinal arrangement between s-CNT and m-CNT. The scale bar indicates 1 μm in (a), and 300 nm in (b).

Figure 3(a) and 3(b) show room temperature drain-current ($I_d$) versus bias-voltage ($V_d$) curves (output characteristics) at different gate-voltages ($V_g$) of representative m-CNT/s-CNT and Pd/s-CNT devices, respectively. We used highly doped silicon as a back-gate and $SiO_2$ as dielectric. Both devices show p-channel transistor behavior along with good current modulation with the gate voltage. The linear $I_d$-$V_d$ curves at a low $V_d$ indicate that, both Pd and m-CNT electrodes form a good interfacial contact with s-CNTs. However, the maximum drain-current (at $V_g$ = -30 V and $V_d$ = -1 V) for Pd/s-CNT devices is only 0.6 μA, whereas it is 12 μA for the m-CNT/s-CNT devices, a value more than an order of magnitude higher.

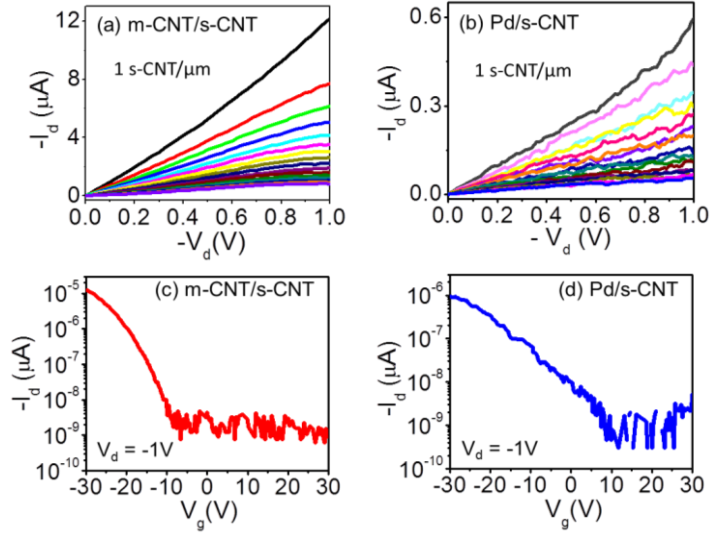

Figure 3(c) and 3(d) show $I_d$-$V_g$ curves (transfer characteristics) at $V_d$ of -1 V for the devices shown in Figure 3(a) and 3(b), respectively. The figure of merits of the device performance such as on-conductance ($G_{on}$= $I_{on}/V_d$), current on-off ratio ($I_{on}/I_{off}$), transconductance ($g_m$ = $dI_d/dV_g$), and linear mobility ($\mu$) were extracted from the $I_d$-$V_g$ curve at $V_d$ = -1 V. We calculated linear mobility ($\mu$) using the standard formula:[15]

**Figure 3.** Output characteristics ($I_d$-$V_d$) for (a) m-CNT/s-CNT and (b) Pd/s-CNT devices at $V_g$ = -30 to -15 V in steps of -1 V (top to bottom). Compared to the Pd/s-CNT device, the m-CNT/s-CNT device shows higher current driving ability. Transfer characteristics (Id-Vg) in semi-log scale for (c) m-CNT/s-CNT and (d) Pd/s-CNT devices at $V_d$ = -1V. The channel length and width of the both m-CNT/s-CNT and Pd/s-CNTs devices are 2 μm and 25 μm respectively.

$$\mu=(L/WC_iV_d)(g_m) \qquad (1)$$

where $C_i$ is the specific capacitance per unit area of aligned array,

$$C_i = D/[C_Q^{-1} + (1/2\pi\varepsilon_0\varepsilon)\ln[\sinh(2\pi t_{ox}D)/\pi Dr]] \qquad (2)$$

$C_Q$ is the quantum capacitance of CNT ($4\times10^{-10}$ F/m), $t_{ox}$ is the oxide thickness (250 nm), $\varepsilon$ is the dielectric constant of $SiO_2$ (3.9), $\varepsilon_0$ is the permittivity of vacuum, $r$ is the average radius of the s-CNTs, and $D$ is the linear density of the s-CNTs in the channel. The $G_{on}$ (at $V_g$ = -30 V), $I_{on}/I_{off}$, $g_m$ and $\mu$ for the Pd/s-CNT device are 0.9 μS, $1.1\times10^3$, 0.4 μS and 5 $cm^2$/Vs, respectively, while for the m-CNT/s-CNT device these values are 15.7 μS, $2.5\times10^4$, 1.5 μS and 51 $cm^2$/Vs respectively. We would like to point out that these values for the m-CNT/s-CNT device are again approximately an order of magnitude higher than that of the control Pd/s-CNT device.

We characterized a total of 26 devices (13 m-CNT/s-CNT and 13 Pd/s-CNT) with the same linear density of 1 s-CNT/μm in the channel. The summary of the device performance is shown in the box plots in Figure 4 (also Table S1). Figure 4(a) shows that the maximum on-conductance for the Pd/s-CNT devices is 10.1 μS with an average of 2.4 μS. In contrast, the maximum and average on-conductance values for the m-CNT/s-CNT devices are 36.5 μS and 19.2 μS respectively, which are significantly higher than those of the Pd/s-CNT devices. Similarly, the maximum and average transconductance of the m-CNT/s-CNT devices, are 2.6 and 1.2 μS, respectively, which are almost one order of magnitude higher than those of the Pd/s-CNT devices, 0.32 and 0.11 μS, respectively, (Figure 4b). Figure 4(c) shows that the maximum mobility of the m-CNT/s-CNT devices is 51 $cm^2$/Vs with an average of 25 $cm^2$/Vs, whereas the maximum mobility of the Pd/s-CNT devices is only 7.8 $cm^2$/Vs with an average of 4 $cm^2$/Vs. Notably, the on-off ratio of the m-CNT/s-CNT devices exceeds that of the Pd/s-CNT devices by two orders of magnitudes. The maximum (average) current on-off ratio of the m-CNT/s-CNT devices is $1.1\times10^6$ ($2.5\times10^5$), whereas it is $1.1\times10^4$ ($2.8\times10^3$) for the Pd/s-CNT devices (Figure 4d). We also compared the performance of our devices with a few recent solution processed short channel (channel length similar to our devices) s-CNT TFT (>98% purity, aligned or random array) devices, as shown in Table 1. From this table we can see that the mobility and current on-off ratio of our Pd/s-CNT TFT (control device) are in agreement with what have been reported in the literature for metal contacted s-CNT TFTs. In contrast, the performance of our m-CNT/s-CNT devices are significantly better than that of our control Pd/s-CNT devices as well as other metal contacted s-CNT devices.

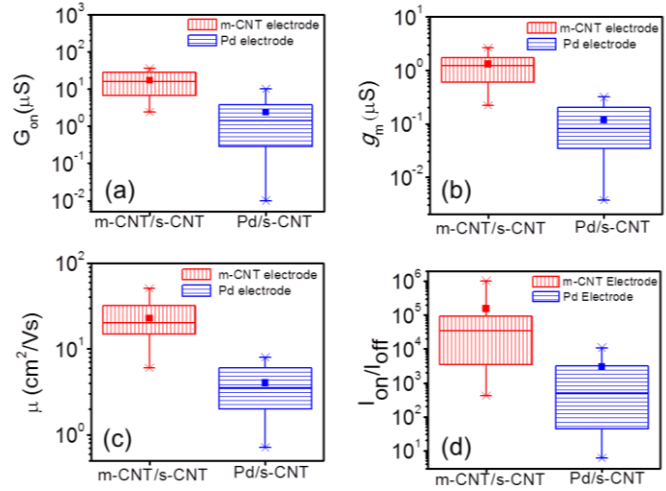

**Figure 4**. Statistics for 13 m-CNT/s-CNT devices (red) and 13 Pd/s-CNT devices (blue). The channel length and width for all devices are 2 μm and 25 μm respectively. (a) on-conductance, (b) transconductance, (c) device mobility, and (d) current on-off ratio. The m-CNT/s-CNTs devices shows significantly better performance than that of the Pd/s-CNT devices.

**Table 1 :** Comparison of a few recent solution-processed short channel s-CNT devices.[a]

| Type of s-CNT (% purity) | Electrode | Channel length (μm) | Channel width (μm) | Mobility ($cm^2$/Vs) | On-off ratio | Reference |
| --- | --- | --- | --- | --- | --- | --- |
| Aligned (99%) | Ti/Pd/Au | 2 | 10 | 10 | $10^4$ | [14] |
| Random (98%) | Ti/Pd | 4 | 100 | 15 | $10^4$ | [16] |
| Random (99%) | Pd/Au | 2 | 200 | 8 | - | [18][b] |
| Random (enriched s-CNT) | Cr/Au | 5 | 60 | 2 | $10^5$ | [26][c] |
| Aligned (99%) | Cr/Pd | 2 | 25 | 7.8 | $1.1\times10^4$ | This work |
| Aligned (99%) | m-CNT | 2 | 25 | 51 | $1.1\times10^6$ | This work |

[a] In this table the devices which have current on-off ratio less than 100 (percolating transport through m-CNT) are not included.
[b] Mobility of 8 $cm^2$/Vs for $L$=2 μm is extrapolated from the mobility versus channel length data.
[c] Percentage of purity is not given.

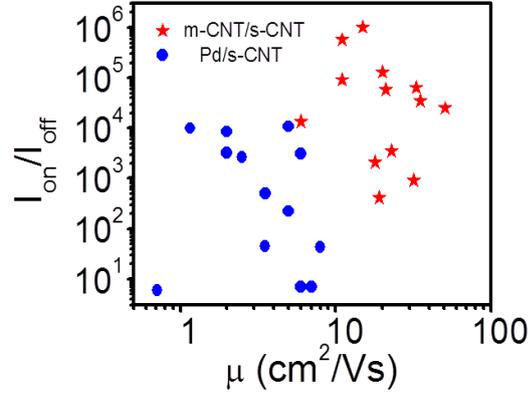

**Figure 5.** Current on-off *versus* mobility of all the m-CNT/s-CNT and Pd/s-CNT devices. Both the mobility and current on-off ratio of the m-CNT/s-CNT devices are higher than that of Pd/s-CNT devices.

In order to be practically applicable in high speed digital electronic devices, transistors should have both higher mobility and a higher current on-off ratio.[2, 18] In Figure 5, we plot the current on-off ratio against mobility for all devices. The star symbols represent the m-CNT contacted devices while the dots represent Pd contacted devices. As shown in Figure 5, the m-CNT contacted devices have consistently higher mobility along with a higher on-off ratio compared to that of the Pd contacted devices. Achieving such a high mobility along with a high current on-off ratio using an m-CNT electrode is very significant for CNT based electronics devices.

The improved performance of a transistor can generally be attributed to several factors including the reduction of traps and impurities in the channel material, improved gate dielectric and improved charge injection due to a lower interface barrier between the metal and semiconductor. Since we used identical channel materials, substrates and gate dielectrics in this work, we believe that the improvement of our m-CNT/s-CNT devices is due to better m-CNT/s-CNT interface compared to the Pd/s-CNT interface. Although it is well-known that Pd makes excellent contact with CVD grown large diameter CNTs, [33] however, our results need to be interpreted in the context that we are dealing with solution processed DEP assembled devices. From the study of solution processed individual CNT transistors it was found that the mobility for individual s-CNT typically varies from 10-1000 cm$^2$/Vs, [32, 34-36] which is significantly lower than the CVD tubes where the typical mobilities are above 1000 cm$^2$/Vs.[37, 38] The inferior performance of the solution processed devices has been attributed to high contact resistance between the CNTs and metal contact as well as defects in the CNTs.[35] Since in the recent studies the highly enriched s-CNT transistors were fabricated using solution processed nanotubes with Pd electrode, our result suggests that a significant improvement of device performance can be achieved by replacing Pd electrode with m-CNT electrode.

In order to clearly understand the reason for improved performance of s-CNT devices using m-CNT electrodes, we calculated the interfacial barriers (Schottky barrier) of our devices from the temperature dependent electronic transport measurements. The $I_d$-$V_d$ curves at $V_g = 0$ V of representative m-CNT/s-CNT and Pd/s-CNT devices in the temperature range of 77 K to 285 K are shown in Figure 6(a) and (b), respectively. These figures show that the drain-currents of both devices decrease with decreasing temperature, indicating a thermally activated transport. From this data, we calculated the Schottky barrier following the analysis of Martel et. al. [39] and others [40-42] by modeling the current using the Arrhenius equation $I_d \sim exp(-\Phi/kT)$, where $k$ is the Boltzmann constant, $T$ is the temperature and $\Phi$ is the Schottky barrier for the s-CNT devices. It was also noted by Martel et. al. and others that $\Phi$ depends on the bias voltage and that the Schottky barrier at zero bias ($\Phi_b$) is the "true" Schottky barrier, which should be calculated by plotting $\Phi(V_d)$ (voltage-dependent Schottky barrier ) as function of the square root of bias voltage ($V_d^{1/2}$) and then extrapolating the linear fit to $V_d^{1/2} = 0V$.[39, 43]

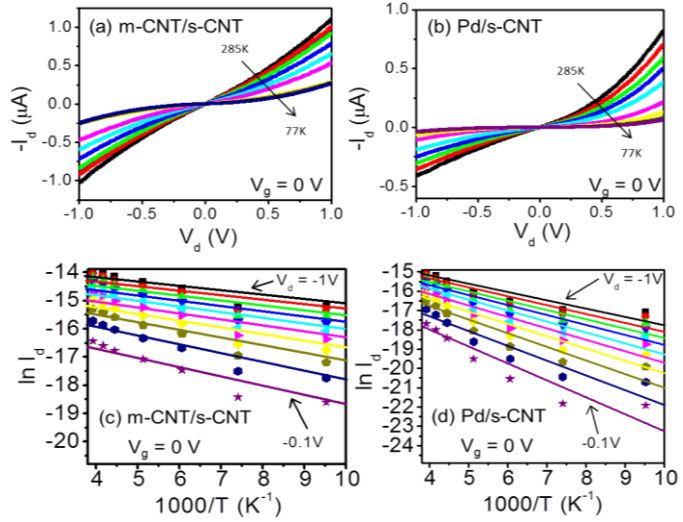

Figure 6(c) and 6(d) show an Arrhenius plot of the current ($lnI_d$ versus $1000/T$) with $V_d$ ranging from -0.1 to -1 V in intervals of -0.1V for m-CNT/s-CNT and Pd/s-CNT devices, respectively. The observed linear relationship between $lnI_d$ and $1/T$ confirms that the transport mechanism in our devices is mainly due to thermally activated charge carriers. From the slopes of the $lnI_d$ versus $1000/T$ curves, we calculated $\Phi(V_d)$ for both m-CNT/s-CNT and Pd/s-CNT devices at different $V_d$. In Figure 7, we plotted $\Phi(V_d)$ as a function of the square root of bias voltage ($V_d^{1/2}$) and determined the Schottky barrier at zero bias ($\Phi_b$) from the intercept of the linear fit of the data. We obtained a value of 24 meV for $\Phi_b$ for the m-CNT/s-CNT device and a value of 89 meV for the Pd/s-CNT device. Interestingly, the Schottky barrier of our aligned arrays Pd/s-CNT devices (average diameter = 1.7 nm) is similar to the Schottky height measured for the Pd contacted individual s-CNT with diameter of 1.7 nm.[44] We have analyzed 3 m-CNT/s-CNT and 3 Pd/s-CNT devices at low temperature (Figure S2). The average $\Phi_b$ of the m-CNT/s-CNT devices is 25 meV, which is significantly lower than the average $\Phi_b$ for Pd/s-CNT devices (83 meV).

**Figure 6**. Temperature dependent (77K - 285K) current-voltage characteristics at $V_g$ = 0V of (a) m-CNT/s-CNT device and (b) Pd/s-CNT device. Arrhenius plot of current at different $V_d$ in the range of -0.1 V to -1V for (c) m-CNT/s-CNT device, and (d) Pd/s-CNT device. The bias voltage dependent Schottky barrier is calculated from the slopes of the $ln I_d$ verse 1000/T curves at different bias voltage.

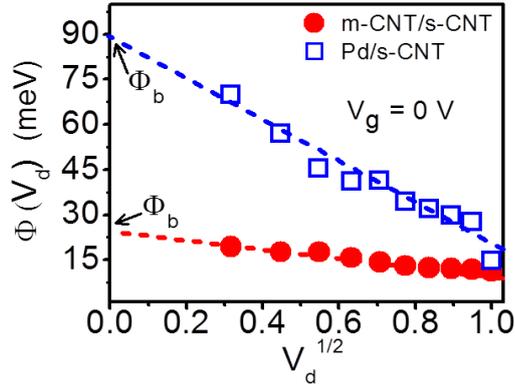

**Figure 7.** Plot of bias voltage dependent Schottky barrier $\Phi$ ($V_d$) a function of square root of bias voltage for m-CNT/s-CNT device (solid circle) and Pd/s-CNT device (open square). The zero bias Schottky barrier ($\Phi_b$) is calculated from the y-intercept.

The possible reason for a small Schottky barrier of our m-CNT/s-CNTs devices can be explained by considering the unique contact configuration between our m-CNTs and s-CNTs at the interfaces. In a previous study, a large Schottky barrier was reported between m-CNT and s-CNT.[45] However, it should be noted that the structure was a cross junction where the CNTs were orthogonal to each other and the contact between them was a point contact. On the other hand, the contact configuration of our devices is completely different: the aligned arrays of s-CNTs are connected between aligned arrays of m-CNTs. This can result in two possible contact configurations: (i) end-contact or (ii) longitudinal side-contact (Figure 8). It has been reported that CNTs have field emission properties due to its one-dimensional character and that the field emission can occur from both the sidewall and tips of the nanotubes.[27, 28, 46-48] It is believed that the field emission properties can enhance charge injection when m-CNTs are used as an electrode material.[30, 31, 49] In the end-contact configuration, the tips of s-CNTs are connected directly to the tips of the m-CNTs (head-to-head connection without any gap) and improved charge injection can occur from the m-CNTs tips due to electric field enhancement at the tips (Figure 8b).[31, 49] When a voltage is applied to m-CNT electrodes, a large local field is generated at the nanotube apex due to their one-dimensional structure which increases charge injection.[31, 49] Moreover, since the m-CNTs electrodes are fabricated by cutting the nanotube arrays, the tips of m-CNTs are open-ended (Figure 1d) and such tips may provide further electric field enhancement and enhance the charge injection.[28, 31] In the side-contact configuration, the m-CNTs and s-CNTs are connected in the longitudinal direction (side walls are parallel) and enhanced charge injection occurs from the side-wall of the m-CNTs (Figure 8c) due to a finite overlap between the tubes, as well as from the tip of the m-CNTs. In contrast, Pd does not have any field emission properties and no electric field enhancement can be expected from Pd electrode, despite there is a finite overlap between Pd and s-CNT. Further theoretical and experimental study will be needed to confirm our hypothesis.

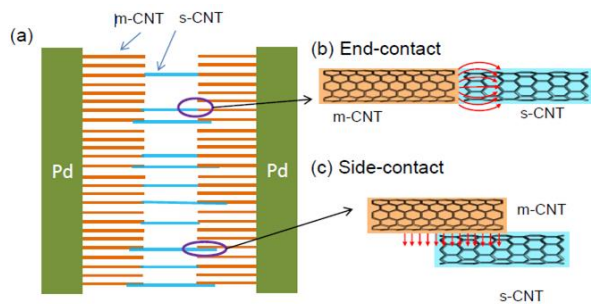

**Figure 8**. (a) Schematic of the aligned arrays s-CNT assembled between aligned array m-CNT electrodes. Two contact configurations are highlighted: end-contact and side-contact. (b) Schematic of end-contact configuration between m-CNT and s-CNT. In the end-contact, the tip of s-CNT is connected directly to the tip of the m-CNT and improved charge injection can occur due to an electric field enhancement at the m-CNT tips. (c) Schematic of side-contact configuration. Improved charge injection can occur from the side-walls of m-CNT due to a finite overlap with s-CNT.

## 3. Conclusion

In summary, we have shown that the performance of the short channel semiconducting enriched aligned array carbon nanotube field effect transistors can be improved significantly by employing metallic carbon nanotube as the electrodes. It has been found that the on-conductance, transconductance, mobility and current on-off ratio of the devices with metallic carbon nanotube electrodes are up to an order of magnitude higher than those of devices fabricated using metal electrodes. From the low temperature electronic transport measurements of our devices, we demonstrated that the improved device performances are due to the lower Schottky barrier of the s-CNT device with metallic carbon nanotube electrodes compared to the metal electrodes. We speculated that the lower Schottky barrier is resulted from the unique one dimensional contact geometry at the interface. This work suggests that, in addition to using semiconducting enriched carbon nanotubes, using metallic carbon nanotube as an electrode can significantly enhance the performance of CNT TFTs.

**Supporting Information**
Supporting Information is available online from the The Royal Society of Chemistry or from the author.

**Acknowledgements**
This work is supported by the U.S. National Science Foundation (NSF) under Grant ECCS 1102228 and ECCS-0748091 (CAREER). We thank Prof. Jing Guo of University of Florida

**The table of contents:**

**High-performance solution processed carbon nanotube (CNT) thin film transistors** are demonstrated using densely aligned arrays of metallic CNTs (m-CNTs) as electrodes, and aligned arrays of semiconducting enriched CNTs (s-CNTs) as channel material. The improved performances are due to a lower charge injection barrier of m-CNT/s-CNT array devices that be attributed to the unique one dimensional contact geometry at the interface where s-CNTs are contacted by m-CNTs via end contact or side contact.

**Keyword**: Semiconducting carbon nanotubes, carbon nanotube electrode, thin film transistor, solution processed, Schottky barrier.

**Title:** High Performance Semiconducting Enriched Carbon Nanotube Thin Film Transistors Using Metallic Carbon Nanotube Electrode

Biddut K. Sarker, Narae Kang, and Saiful I. Khondaker*

ToC figure

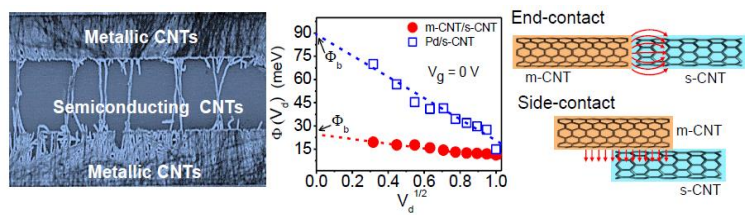

# Supporting Information

**High Performance Semiconducting Enriched Carbon Nanotube Thin Film Transistors Using Metallic Carbon Nanotube Electrode**

Biddut K. Sarker, Narae Kang, and Saiful I. Khondaker*

1. **Electrical characterization for aligned array CNTs.**

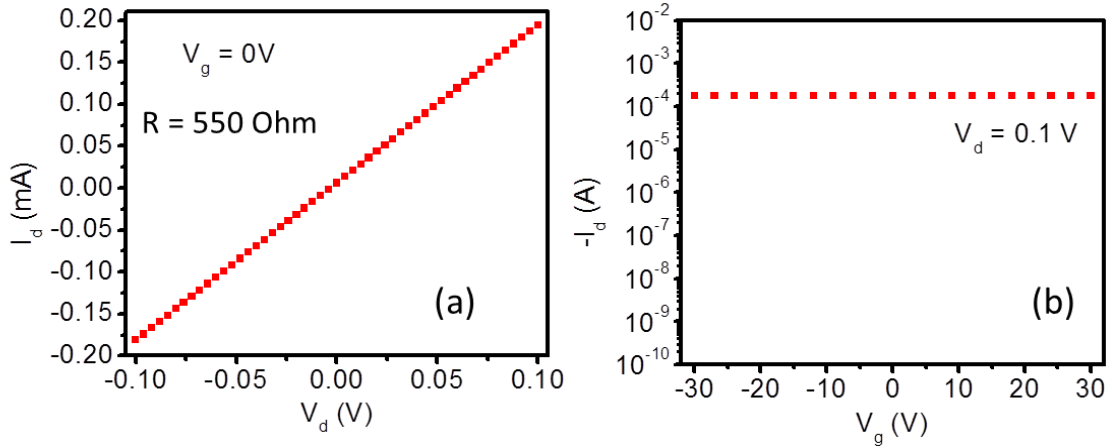

**Figure S1**: (a) Current-voltage ($I_d$-$V_d$) characteristics of the highly dense aligned array single walled carbon nanotubes (CNTs) at zero gate voltage ($V_g$). The resistance of this array is 550 Ω and corresponding sheet resistance is 2.75 KΩ. (b) current-gate voltage ($I_d$-$V_g$) characteristics of the highly dense CNT aligned array at source-drain voltage ($V_d$) of = -0.1V. As we see from this figure that the current of the arrays does not charge with gate voltage. This clearly indicating that the CNT arrays used for fabricating CNT electrodes in this study are metallic in nature. We have tested ($I_d$-$V_d$) and ($I_d$-$V_g$) for a series of densely CNT aligned arrays and all the arrays have shown low resistance and metallic nature.

## 2. Temperature-dependent transport properties of another m-CNT/s-CNT device.

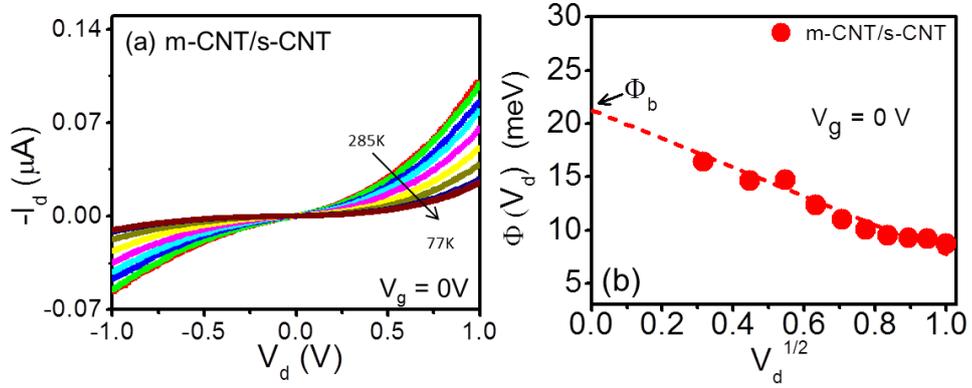

**Figure S2**: Temperature dependent charge transport properties of another m-CNT/s-CNT device. (a) Current-voltage characteristics at $V_g = 0$ for different temperatures in the range of 285 K - 77 K (b) Plot of bias voltage dependent Schottky barrier $\Phi$ ($V_d$) a function of square root of bias voltage for m-CNT/s-CNT device. The zero bias Schottky barrier $\Phi_b = 22$ meV is calculated from the y-intercept.

## 3. Summary of all measured devices at room temperature :

**Table S1**: The on-conductance ($G_{on}$), transconductance ($g_m$), mobility ($\mu$), and current on-off ratio ($I_{on}/I_{off}$) for the devices with m-CNT/s-CNT and Pd/s-CNT.

|  |  | $G_{on}$ (µS) | $g_m$ (µS) | $\mu$ (cm$^2$/Vs) | $I_{on}/I_{off}$ |
|---|---|---|---|---|---|
| m-CNT/s-CNT | max. | 36.50 | 2.63 | 51 | $1.10 \times 10^6$ |
|  | avg. | 19.21 | 1.19 | 25 | $2.45 \times 10^5$ |
| Pd/s-CNT | max. | 10.10 | 0.32 | 8 | $1.1 \times 10^4$ |
|  | avg. | 2.41 | 0.11 | 4 | $2.80 \times 10^3$ |